\documentclass[useAMS,usenatbib]{mnras}
\usepackage{amsmath}
\usepackage{times}
\usepackage{graphicx}
\title[Giant outburst of SXP 59]
{{\it NuSTAR} and {\it XMM--Newton} observations of SXP~59 during its 2017 giant outburst}

\author[Weng et al.]{Shan-Shan Weng$^{1}$\thanks{E-mail: wengss@njnu.edu.cn}, Ming-Yu Ge$^{2}$\thanks{E-mail: gemy@ihep.ac.cn}, Hai-Hui Zhao$^{1}$\thanks{E-mail: zhaohh@njnu.edu.cn} \\
$^1$\,Department of Physics and Institute of Theoretical Physics,
Nanjing Normal University, Nanjing 210023, China \\
$^2$\,Key Laboratory of Particle Astrophysics, Institute of High Energy
Physics, Chinese Academy of Sciences, Beijing 100049, China \\
}

\date{}

\begin{document} \maketitle

\begin{abstract}
The Be X-ray pulsar (BeXRP) SXP~59 underwent a giant outburst in 2017 with a peak X-ray luminosity of $1.1\times10^{38}$ erg~s$^{-1}$. We report on the X-ray behaviour of SXP~59 with the {\it XMM--Newton} and {\it NuSTAR} observations collected at the outburst peak, decay, and the low luminosity states. The pulse profiles are energy dependent, the pulse fraction increases with the photon energy and saturates at $\sim$ 65\% above 10 keV. It is difficult to constrain the change in the geometry of emitting region with the limited data. Nevertheless, because the pulse shape generally has a double-peaked profile at high luminosity and a single peak profile at low luminosity, we prefer the scenario that the source transited from the super-critical state to the sub-critical regime. This result would further imply that the neutron star (NS) in SXP~59 has a typical magnetic field. We confirm that the soft excess revealed below 2 keV is dominated by a cool thermal component. On the other hand, the {\it NuSTAR} spectra can be described as a combination of the non-thermal component from the accretion column, a hot blackbody emission, and an iron emission line. The temperature of the hot thermal component decreases with time, while its size remains constant ($R \sim 0.6$ km). The existence of the hot blackbody at high luminosity cannot be explained with the present accretion theories for BeXRPs. It means that either more sophisticated spectral models are required to describe the X-ray spectra of luminous BeXRPs, or there is non-dipole magnetic field close to the NS surface.
\end{abstract}


\section{Introduction}

In general, Be X-ray binary (BeXRB) consists of a young neutron star (NS) orbiting a Be type star. Most BeXRBs are transient in X-ray, and their variabilities are often classified into two types of outbursts \citep[see ][for reviews]{bildsten97, reig11}. Type I outbursts are less energetic ($L_{\rm peak} < 10^{37}$ erg~s$^{-1}$) and occur regularly as the enhancement of accretion during the periastron passage. On the other hand, type II outbursts are rare and not fixed to the orbital phase. Their X-ray luminosity can exceed the Eddington luminosity for a NS. In particular, the peak X-ray luminosities of the 2016-17 outburst of SMC X-3 \citep[e.g. ][]{weng17, zhao18} and 2017-18 outburst of Swift~J0243.6+6124 \citep{doroshenko18, wilson18, tao19} are beyond $10^{39}$ erg~s$^{-1}$, that is the threshold of ultraluminous X-ray sources \citep[ULXs, ][]{kaaret17}.

Hundreds of high-mass X-ray binaries (HXMBs) have been detected in the Galaxy and the Magellanic Clouds, and more than half of them are BeXRBs \citep{liu05, liu06}. In addition to the short distance \citep[62.1 kpc; ][]{hilditch05, graczyk14, scowcroft16}, the Small Magellanic Cloud (SMC) has a large star formation rate \citep[150 times of the Galaxy; ][]{harris04} and low interstellar absorption \citep{zaritsky02, willingale13}.  It thus provides an ideal and large sample of BeXBRs for a detailed study in multibands \citep[e.g. ][]{rajoelimanana11, bird12, coe15}. Historically, many X-ray observatories (e.g. {\it ROSAT}, {\it RXTE}, {\it Chandra}, {\it XMM--Newton}) had spent a lot of time to survey HXMBs in the SMC \citep[e.g. ][]{kahabka99, galache08, sturm13, haberl16, yang17}. Since 2016 June, the Neil Gehrels {\it Swift} Observatory has started a high cadence shallow (with a typical exposure of 60 s) survey of the SMC in order to monitor X-ray variabilities of BeXRBs by taking its advantage of rapid slewing. During the first year operation, the {\it Swift} SMC survey (S-CUBED) successfully detected the type II outbursts from SMC~X-3, SXP~59, and SXP~6.85 \citep[see ][for more details]{kennea18}.

SXP~59 was identified as an X-ray pulsar ($P = 59.0\pm0.2$ s) in 1998 due to its outburst, while the pulsations with the same period were also revealed in the {\it ROSAT} archive data \citep{marshall98}. The orbital period of $\sim 122.1$ d was reported with both the {\it RXTE} observations \citep{galache08} and the OGLE {\it I}-band light curves \citep{bird12}. S-CUBED detected the onset of giant outburst from SXP~59 on 2017 March 30 \citep{kennea17}, and the {\it Swift} TOO observations were triggered to follow the outburst. The source reached a peak luminosity on 2017 April 7 ($\sim 4.6\times10^{37}$ erg~s$^{-1}$ in 0.5--10 keV), then exponentially declined with an time-scale of $\sim 15.9$ d, and returned to the pre-outburst flux level on 2017 June 6 \citep{kennea18}. Investigating the {\it XMM--Newton} TOO observation performed around the peak of outburst, \citet{la18} revealed a soft excess below 2 keV in addition to the primary power-law component. Since the double-peaked pulse profile detected at the high-luminosity level, they also speculated that the source was at the super-critical state having a fan-beam emission geometry.

In this paper, we carry out a detailed analysis on the high-quality data obtained from the {\it XMM--Newton} and another three {\it NuSTAR} observations executed at different flux levels to explore the spectral evolution of SXP~59 during its 2017 giant outburst. Section 2 describes the observations together with the data analysis, and summarizes our results.  We discuss the physical implications of these results in Section 3.

\section{Data Analysis}
\subsection{X-ray observations}

\begin{figure}
\begin{center}
\includegraphics[width=8cm]{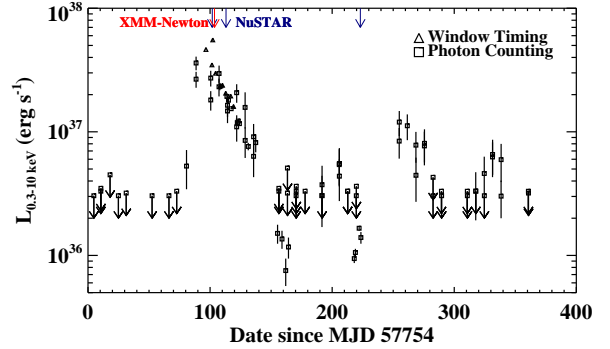}
\end{center}
\caption{{\it Swift}/XRT light curve of SXP 59 since 2017 Jan 1 (MJD~57754). 2 $\sigma$ upper limits for non-detections are shown with black arrows. The red and blue arrows label the {\it XMM--Newton} and the {\it NuSTAR} observations, respectively. \label{lc}}
\end{figure}

In 2017, the Neil Gehrels {\it Swift} Observatory carried out 92 observations on SXP~59, including 65 S-CUBED observations. In order to avoid the pile-up effect, the TOO observations around the outburst peak were executed with the window timing (WT) mode instead of the photon counting (PC) mode. The {\it Swift}/XRT data are processed with the packages and tools available in \textsc{heasoft} 6.24. The software \texttt{xrtpipeline} is used with standard quality cuts for the initial event cleaning. We extract the source light curves in 0.3--10 keV from a circle of 15 pixels centred at the source position, and the background light curves from an annulus region with the radii of 15 and 30 pixels. The source light curves are corrected for the telescope vignetting and point spread function losses with the task \texttt{xrtlccorr}, and then are subtracted by the scaled background count rate to generate the net light curves (Figure \ref{lc}). When the source was in quiescence state, it can hardly be detected by the S-CUBED observations due to their short exposures. Following the work in \cite{kennea18}, we adopt five counts as the threshold of detection, and calculate the upper limits for non-detections. Based on the following {\it XMM--Newton} spectral fitting results, we convert the count rate to the flux, and hence the luminosity assuming a distance of 62.1 kpc. The derived count rate to luminosity ratios for the WT and the PC modes are of 1~count~s$^{-1}$ $\sim 3.55\times10^{37}$ erg~s$^{-1}$ and $\sim 3.77\times10^{37}$ erg~s$^{-1}$, respectively. As can be seen in Figure \ref{lc}, the giant outburst lasted for about two months with a fast-rise near exponential decay profile. These results are consistent with those reported in \cite{kennea18} (Figure 13 in their paper).

In this work, we analyze the {\it XMM--Newton} observation performed on 2017 April 14, which was free of the background contamination. The data collected with the {\it XMM--Newton} EPIC instrument are reduced using the Science Analysis System software (\textsc{sas}) version 14.0.0. Both the pn and MOS data were taken in small window mode in order to minimize the pile-up effect. We exclude all events at the edge of CCD and from bad pixels by setting FLAG=0, select the pn events with PATTERN in the 0-4 range, and the MOS data with PATTERN$\leq$12. The source photons are extracted from a circle aperture with a radius of 30 arcsec, and the background is taken from the same CCD chip as the source within a circle of radius 50 arcsec.

{\it NuSTAR} is the first direct-imaging hard X-ray telescope, consisting of two focusing instruments and two focal plane modules, i.e. Focal Plane Modules A and B \citep[hereafter FPMA and FPMB; ][]{harrison13}. There are three {\it NuSTAR} observations carried out at the outburst peak, decay, and the low luminosity states, respectively (Table \ref{log} and Figure \ref{lc}). The source events are extracted from circular region with the radius of 60--100 arcsec, depending on the count rates. Meanwhile, the background photons are extracted from the source-free region with a radius of 120{\arcsec}. These data are processed with the task \texttt{nupipeline}, the spectra and the light curves are produced with the command \texttt{nuproducts}. It is worth to note that, the first {\it NuSTAR} data were made 1--2 d before the {\it XMM--Newton} observation. That is, these two observations are quasi-simultaneous.

\begin{table*}\small
\centering \begin{tabular}{clcccccc} \hline
Obs Date & Observatory & ObsID & Exposure &  Epoch  & $\nu$  & $\dot{\nu}$ \\
   &      &     & (ksec)   & (MJD-57850) & ($10^{-2}$ Hz) & ($10^{-11}$ Hz s$^{-1}$) \\
\hline

2017 Apr 14     & {\it XMM--Newton}  & 0740071301 & 14 & 7.617424 & 1.69633(4) & ... \\ \hline
2017 Apr 12--13     & {\it NuSTAR}  & 30361001002 & 70 & 5.786459 & 1.69570(2) & 3.4(4)  \\
2017 Apr 24--26     & {\it NuSTAR}  & 50311001002 & 153& 5.786930 &1.69669(7) & 2.02(7) \\
2017 Aug 12--13     & {\it NuSTAR}  & 50311001004 & 82 & 127.093712 & 1.700745(7) & ... \\
\hline
\end{tabular}
\caption{Log of {\it XMM--Newton} and {\it NuSTAR} observations. \label{log}}
\end{table*}

\subsection{Spectral analysis}

Both {\it NuSTAR} and {\it XMM--Newton} spectra are fitted by empirical models most often used in the literature with the \textsc{heasoft} X-ray spectral fitting package \textsc{xspec} \citep{arnaud96}. All models in this paper also include the interstellar absorption (\texttt{tbabs} in \textsc{xspec}). The {\it NuSTAR} spectra are grouped with \texttt{grppha} to ensure at least 30 counts per bin. The FPMA and the FPMB spectra are fitted simultaneously, with a constant multiplicative factor to compensate for calibration differences. The FPMA constant is fixed at unity, whilst that for the FPMB is allowed to vary, with the yielded values in the range of 1.00--1.06.

\begin{figure}
\begin{center}
\includegraphics[width=8cm]{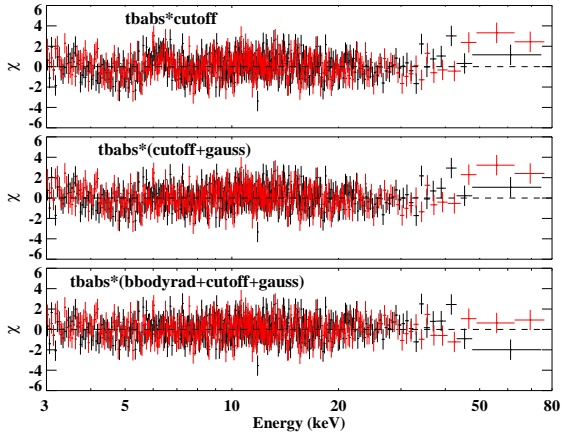}
\end{center}
\caption{Spectra of the first {\it NuSTAR} observation are fitted the models of \texttt{tbabs*cutoffpl}, \texttt{tbabs*(cutoffpl+gaussian)}, and \texttt{tbabs*(bbodyrad+cutoffpl+gaussian)}, respectively. Panels from top to bottom show the corresponding fit residuals.  \label{nu_chi}}
\end{figure}

Because the {\it NuSTAR} and {\it XMM--Newton} data are operated in different energy ranges (3--79 keV and 0.5--10 keV), some emission component might be caught by only one of them. Thus, we firstly decompose the spectral components with {\it NuSTAR} and {\it XMM--Newton} spectra separately, and aim to achieve a common model for the broadband spectra. We begin by fitting the cut-off power-law component to the first {\it NuSTAR} observation. The derived reduced chi-square is of $1.09$ ($\chi^{2}/dof = 1139.0/1045$), the iron line feature and residuals at low and high-energy bands are displayed in the top panel of Figure \ref{nu_chi}. Thus, a Gaussian line at $\sim 6.3$ keV is added to account for the iron line component. The fitting is further significantly improved with an additional hot thermal component ($kT \sim 4.1$ keV). The reduced chi-square decreases from $\chi^{2}/dof = 1105.7/1042$ to $\chi^{2}/dof = 1047.8/1040$, and the fit residuals become flat in the whole energy band (bottom panel of Figure \ref{nu_chi}). The same situation occurs for the {\it NuSTAR} data obtained at the outburst decay phase. Alternatively, the blackbody component is required with a confidence level of 98\% according to $F$-test, but the iron line is too weak to be detected in the last {\it NuSTAR} observation. We, therefore, suggest that the {\it NuSTAR} spectra can be described as a combination of a hot blackbody and a cut-off power-law component, and the iron line emission is required at the high luminosity state (Table \ref{spectra}).

\begin{figure}
\begin{center}
\includegraphics[width=8cm]{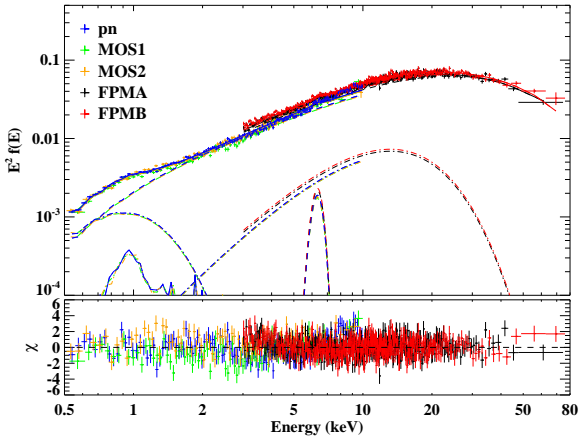}
\end{center}
\caption{A joint fit is applied to the first {\it NuSTAR} and {\it XMM--Newton} data with the model of \texttt{tbabs*(apec+bbodyrad+bbodyrad+cutoffpl+gaussian)}.  Unfolded spectra  (upper panel) and fit residuals (bottom panel) are plotted. Dash--dotted lines mark two blackbody components.  \label{joint_fit}}
\end{figure}

For the {\it XMM--Newton} data, we generate the spectral response files with the \textsc{sas} tasks \texttt{rmfgen} and \texttt{arfgen}, and rebin the spectra by using the task \texttt{specgroup} to have at least 20 counts per bin to enable the use of chi-square statistics and not to oversample the instrument energy resolution by more than a factor of three. \cite{la18} carried out a detailed analysis on the {\it XMM--Newton} data, and concluded that the continuum spectrum was dominated by the power-law component, and displayed a soft excess below 2 keV. The latter feature was further described with the sum of a cool blackbody and a hot thermal plasma component. Here, we fit the pn and MOS1/2 data simultaneously and confirm that all three components are required by the data. Adopting the same model as used in \cite{la18} [\texttt{tbabs*(apec+bbodyrad+powerlaw+gaussian)} in \textsc{xspec}] with the metal abundance of the APEC component fixed to 0.2 $Z_{\odot}$,  we obtain the similar values for all parameters: $nH = 0.07_{-0.02}^{+0.03}\times10^{22}$ cm$^{-2}$, $kT_{\rm apec} = 0.96_{-0.09}^{+0.08}$ keV, Norm$_{\rm apec} = 6.4_{-2.1}^{+2.0}$, $kT_{\rm BB}^{\rm low} = 0.22_{-0.02}^{+0.03}$ keV, Norm$_{\rm BB}^{\rm low} = 86.3_{-47.8}^{+90.4}$, $\Gamma = 0.73_{-0.02}^{+0.02}$, Norm$_{\rm PL} = 2.66_{-0.09}^{+0.09}\times10^{-3}$, $E_{\rm Gau} = 6.34 _{-0.25}^{+0.24}$ keV, $\sigma = 0.34_{-0.21}^{+0.28}$ keV, Norm$_{\rm Gau} = 3.25_{-1.89}^{+2.31}\times10^{-5}$, and $\chi^{2}/dof = 575.2/432$). But because the peak emission of the hot blackbody component needed by the {\it NuSTAR} data is beyond 10 keV (Figure \ref{joint_fit}), its parameters cannot be constrained with the {\it XMM--Newton} spectra alone.

\begin{figure}
\begin{center}
\includegraphics[width=8cm]{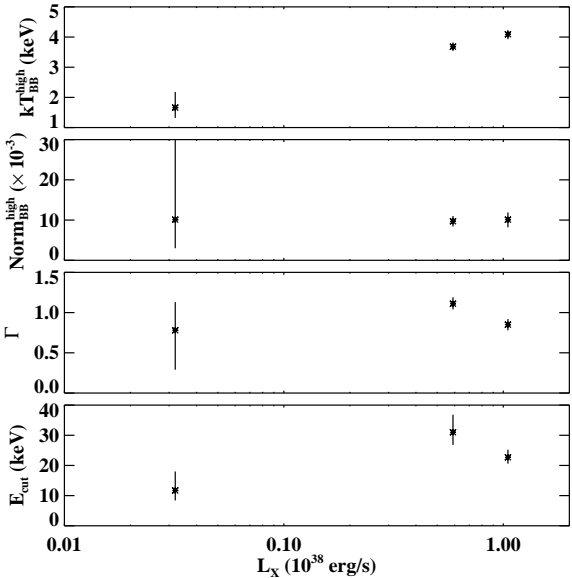}
\end{center}
\caption{Spectral parameters vary as a function of X-ray luminosity (Table \ref{spectra}).}  \label{nu_fit}
\end{figure}

Since the separation of first {\it NuSTAR} observation and the {\it XMM--Newton} observation is less than 2 d, we also try to fit them together with the common model of \texttt{tbabs*(apec+bbodyrad+bbodyrad+cutoffpl+gaussian)} as discussed above. The  multiplicative constant for pn data is frozen at unity, and those for MOS1/2 and FPMA/B are allowed to vary. The derived constant factors for MOS1/2, FPMA, and FPMB are $0.97\pm0.01$, $1.15\pm0.01$, and $1.22\pm0.01$, respectively. The unfolded spectra are plotted in Figure \ref{joint_fit}, and the spectral parameters are presented in Table \ref{spectra} and Figure \ref{nu_fit}. The hot thermal plasma, the two blackbody and the non-thermal components contribute the X-ray emissions (in 0.5--79 keV) of $\sim$ $2.4\times10^{35}$ erg~s$^{-1}$, $1.3\times10^{36}$ erg~s$^{-1}$, $6.9\times10^{36}$ erg~s$^{-1}$, and $1.02\times10^{38}$ erg~s$^{-1}$, respectively. There is no obvious evidence of cyclotron absorption line feature in either the {\it XMM--Newton} nor the {\it NuSTAR} data. Finally, we would caution that the small discrepancy between the joint-fitting results and those obtained from fitting the {\it XMM--Newton} and the {\it NuSTAR} data alone could be due to the calibration differences and the spectral evolution within 2 d.

\begin{table}\scriptsize
\centering \begin{tabular}{l c c c c}
\hline Parameters & {\it XMM}+{\it NuSTAR} & {\it NuSTAR} & {\it NuSTAR} & {\it NuSTAR} \\
%\hline
                  &  Apr 12--14 &  Apr 12--13 & Apr 24--26  & Aug 12--13   \\
\hline
$nH$ ($10^{22}$ cm$^{-2}$) &  $0.10_{-0.03}^{+0.03}$ & 0.10 (fixed) & 0.10 (fixed)  &  0.10 (fixed)  \\
$kT_{\rm apec}$ (keV) &  $1.00_{-0.12}^{+0.16}$ &   --     &   --      &       --   \\
Norm$_{\rm apec}$ ($\times 10^{-4}$)&  $5.3_{-2.4}^{+2.2}$  &   --     &   --      &       ...   \\

$kT_{\rm BB}^{\rm low}$ (keV) &  $0.19_{-0.01}^{+0.01}$ &   --     &   --      &       ...   \\
Norm$_{\rm BB}^{\rm low}$ &  $226.5_{-109.2}^{+184.8}$  &   --     &   --      &       ...   \\

$kT_{\rm BB}^{\rm high}$ (keV) &  $3.36_{-0.25}^{+0.21}$ & $4.09_{-0.16}^{+0.14}$ &$3.68_{-0.14}^{+0.14}$&       $1.66_{-0.34}^{+0.52}$ \\

Norm$_{\rm BB}^{\rm high}$ ($\times 10^{-3}$) &  $9.5_{-2.2}^{+2.2}$  & $10.1_{-1.9}^{+1.8}$ & $9.7_{-1.3}^{+1.3}$ &    $10.1_{-7.1}^{+21.6}$  \\

$\Gamma$  &  $0.64_{-0.04}^{+0.05}$ &  $0.85_{-0.07}^{+0.07}$ &$1.11_{-0.07}^{+0.08}$&       $0.78_{-0.49}^{+0.35}$ \\

$E_{\rm cut}$ (keV) &  $17.8_{-1.0}^{+1.2}$ & $22.7_{-2.1}^{+2.5}$ & $31.0_{-4.2}^{+5.8}$ &       $11.7_{-3.3}^{+6.3}$ \\

Norm$_{\rm cut-off}$ ($\times 10^{-3}$) & $2.8_{-0.1}^{+0.1}$ & $4.2_{-0.3}^{+0.3}$ & $3.7_{-0.3}^{+0.3}$&       $0.25_{-0.14}^{+0.13}$  \\

$E_{\rm Gau}$ (keV) &  $6.30_{-0.09}^{+0.09}$  &  $6.31_{-0.10}^{+0.10}$ & $6.26_{-0.15}^{+0.19}$ & --  \\

$\sigma$ (keV) &  $0.33_{-0.09}^{+0.10}$  &  $0.31_{-0.10}^{+0.11}$ &   $0.55_{-0.19}^{+0.34}$  &     -- \\

Norm$_{\rm gauss}$ ($\times 10^{-5}$)& $3.9_{-0.9}^{+1.0}$ & $4.6_{-1.2}^{+1.3}$ & $3.8_{-1.2}^{+1.9}$ & --\\

$L_{\rm X}$ ($10^{38}$ erg s$^{-1}$)  &  1.11$^{a}$  & 1.05$^{b}$ & 0.59$^{b}$ &       0.032$^{b}$  \\

$\chi^{2}/dof$  &  1806.2/1476 &  1047.8/1040 &  1119.3/975 &     408.0/403 \\
\hline
\end{tabular}
\caption{Spectra are fitted with the model of \texttt{tbabs*(apec+bbodyrad+bbodyrad+cutoffpl+gaussian)}. $^{a}$Unabsorbed luminosity is calculated in 0.5--79 keV by assuming a distance of 62.1 kpc. $^{b}$Unabsorbed luminosity is calculated in 3--79 keV.  All errors are in 90\% confidence level.}
\label{spectra}
\end{table}

In order to verify the existence of the blackbody components, we also try to use other commonly used models to describe the non-thermal X-ray continuum \citep[e.g. ][]{coburn02, west17}, such as the negative and positive exponential cut-off \citep[the so-called NPEX model, ][]{mihara98}, the Fermi Dirac cut-off \citep[the so-called FDCut model, ][]{tanaka86}, and the high-energy cut-off power-law models (\texttt{highecut*powerlaw} in XSPEC). These models predict that the spectra having a power-law profile below 10 keV and rolling off in different ways at high-energy band. The soft excess revealed below 2 keV is not sensitive to the adopted non-thermal models. On the other hand, the temperature of hot thermal component does not change much while its emitting size could vary by a factor of $< 3$ when different continuum models are used. That is much smaller than the radius of a NS. Alternatively, a more physical model, CompTT, is also used to fit the spectra resulting in the similar parameter values for the thermal component, but obtain a worse fit. Note that, compared to the cut-off power-law model, these models have more parameters, which sometimes are difficult to be constrained. In sum, we suggest that the spectral parameters yielded by the cut-off power-law model are reliable and can be better constrained.

\subsection{Pulse profiles and pulse fractions}
The 0.3--12 keV source events are extracted from {\it XMM--Newton} EPIC data and are barycentrically corrected with the command \texttt{barycen}. Meanwhile, for the {\it NuSTAR} data, the source events are extracted in 3--79 keV for the period calculation.
For each observation, an accurate template profile with 50 phase bins is created by folding the whole event data. Then we divide one observation into several segments having equal exposure (4000 s), and derive the pulse times of arrivals (TOAs) of the pulsar by comparing the template profile with the one from each segment, as detailed in the following: (1) search for the best spin frequency using the Pearson $\chi^2$ method; (2) fold the pulse profile with the starting time of the observation as the reference epoch; (3) Calculate the phase shift using the cross-correlation between the pulse profile and the template profile, which represents the TOA of each observation. Finally, we determine the rotation frequencies and their derivatives for each observation by fitting the TOAs with TEMPO2 \citep[][ Table \ref{log}]{hobbs06}.

\begin{figure}
\begin{center}
\includegraphics[width=8cm]{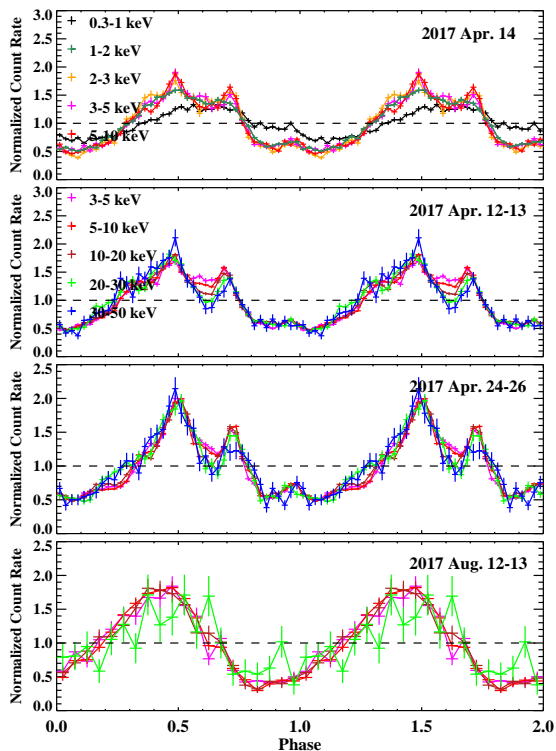}
\end{center}
\caption{Evolution of energy-dependent pulse profiles.  \label{pp}}
\end{figure}

\begin{figure}
\begin{center}
\includegraphics[width=8cm]{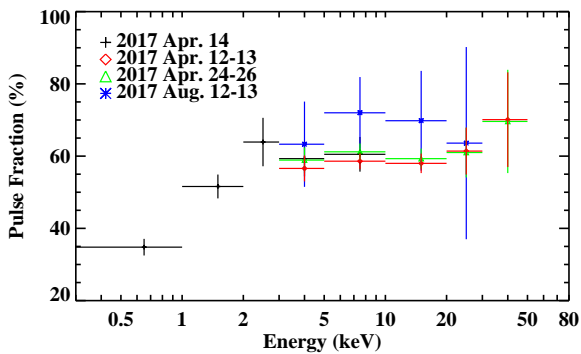}
\end{center}
\caption{Energy-dependent pulse fraction.  \label{pf}}
\end{figure}

We also produce the light curves in time resolution of 0.1 s. The barycentric corrected light curves are folded over the best-fitting period, and the pulse fractions are calculated as $PF = (C_{\rm max}-C_{\rm min})/(C_{\rm max}+C_{\rm min})$, where $C_{\rm max}$, $C_{\rm min}$ are the maximum and the minimum count rates of the profile. The evolutions of the pulse profile and pulse fraction are plotted in Figures \ref{pp} and \ref{pf}, respectively. However, we cannot investigate the pulse modulation above 50 keV (30 keV) for the first two (the last) {\it NuSTAR} observations owing to the low count rate.

\subsection{Results}
Our results are summarized  as follows: (I) During the 2017 giant outburst, SXP~59 reached a peak luminosity of $\sim 1.1\times10^{38}$ erg~s$^{-1}$, that is 60\% Eddington luminosity of a NS. (II) Investigating the {\it XMM--Newton} data, we confirm that the soft excess reported by \cite{la18} consists of a cool thermal component ($kT_{\rm BB} \sim 0.2$ keV) with a size of $10^{2}$ km and a hot thermal plasma . (III) The hard X-ray spectra ($> 3$ keV) are modelled by three components: a hot blackbody component ($kT_{\rm BB} \sim 1.5-4$ keV), a non-thermal component, and an iron emission line. The temperature of blackbody decreases with time, while its normalization remains constant ($\sim 0.01$, Figure \ref{nu_fit}), corresponding to a size of $R \sim 0.62$ km. (IV) The pulse profiles given by the first two {\it NuSTAR} data are energy dependent and have two narrow peaks at phase of $\sim$ 0.5 and 0.7. Alternatively, the pulse shape at the low luminosity state has a single peak profile. (V) The pulse fraction increases with the photon energy and saturates at 65\% above 10 keV for all three {\it NuSTAR} observations.

\section{Discussions AND Conclusions}

The accretion geometry in BeXRBs is mainly governed by the NS magnetic field strength ($B$) and the accretion rate \citep{basko76, riffert88, kraus95, becker12, mushtukov18}. For the case of low accretion rate, the falling material is funnelled by the magnetic field to small regions around the polar caps of NSs (i.e. hot spots). The X-ray flux is mainly contributed by the thermal component from the hot spots with a temperature of $> 1$ keV and a small radius of $< 1$ km, e.g. SAX~J2103.5+4545 \citep{inam04}, 1A 0535+262 \citep{mukherjee05}, RX~J1037.5-5647 \citep{la09}.  Theoretically, the size of the hot spot increases with luminosity \citep{lamb73, frank02, mushtukov15}. When the interaction between the thermal photons and the falling material (bulk motion Comptonization) is non-negligible, the observed spectrum would be deviated from the blackbody form, but can be fitted the CompTT model in \textsc{xspec} \citep[e.g. ][]{doroshenko10, tsygankov19}. As the accretion is larger than the critical value, the accretion column is formed and blocks the sight of hot spots. That is, the X-ray flux is dominated by the non-thermal component from the accretion column, and no emission from hot spots is expected.

The change of beam pattern (i.e. the existence of accretion column or hot spots at the stellar surface, the so-called fan beam and pencil beam) results in different pulse profiles. It has been observed in several giant outbursts of BeXRBs that, the pulse shapes transit from double peaks at high luminosity to single peak at low luminosity, and the pulse fraction increases with energy, e.g. 1A~0535+262 \citep{bildsten97}, SMC~X-3 \citep{weng17, zhao18}, and Swift~J0243.6+6124 \citep{tsygankov18, wilson18}. Such evolution sequence can be interpreted as the different radiation beam patterns working in the super-critical and sub-critical accretion regimes \citep{basko76, becker12, mushtukov15}. Taking account of the exact Compton scattering cross section in a strong magnetic field, \cite{mushtukov15} argued that the critical luminosity was not a monotonic function of $B$, and it reached a minimum of a few $10^{36}$ erg~s$^{-1}$ when the cyclotron energy was about 10--20 keV (fig. 5 in their paper).

SXP~59 entered into a type II outburst in 2017 and became one of the brightest BeXRBs with a peak X-ray luminosity of $1.1\times10^{38}$ erg~s$^{-1}$. Investigating the {\it XMM--Newton} and {\it NuSTAR} observations executed at different flux levels, we find that the pulse profiles evolve both with the photon energy and the X-ray luminosity (Figure \ref{pp}). In general, the pulse profiles above 2 keV exhibit two narrow peaks at the high luminosity, and turn into a single peak in the last {\it NuSTAR} data. Although it is difficult to constrain changes in the geometry of emitting region with the data presented in this work, our results are in favor of the scenario that the source transited from the super-critical state to the sub-critical state as observed in 1A~0535+262 and SMC X-3. The critical luminosity is of $3.2 \times 10^{36}~{\rm erg~s^{-1}} < L_{\rm crit} < 5.9 \times 10^{37}~{\rm erg~s^{-1}}$., that is a typical value for a Be X-ray pulsar \citep[e.g. ][]{becker12, mushtukov15}. It might further suggest a typical magnetic field ($\sim 10^{12}-10^{13}$ G) for the NS in SXP~59, although we cannot put tight constraint at current stage. It worth to note that the cyclotron absorption line feature is the direct evidence for the NS magnetic field; however, it could be transient and too weak to be detected. For instance, the bursting pulsar, GRO~J1744-28 was discovered in 1995 \citep{fishman95} and since then has been observed frequently by X-ray missions (e.g. {\it BeppoSAX}, {\it RXTE}, {\it XMM--Newton}, {\it Chandra}, and {\it NuSTAR}); but the weak absorption feature at $\sim$ 4.5 keV was detected only recently \citep{dai15, doroshenko15}. Therefore, the absence of cyclotron absorption line is not in contradiction with a typical magnetic field for SXP~59.

It was reported that, a cool thermal emission ($kT_{\rm BB} \sim 0.2$ keV) with a large emission area emerged in the {\it XMM--Newton} data of SXP~59 \citep{la18}. The spectral modelling parameters along with the significantly small pulse fraction detected below 1 keV  ($< 35\%$, Figure \ref{pp}), are in favor of the scenario that the central hard X-rays are reprocessed by the inner region of the accretion disc \citep{hickox04, la18}. The X-ray continuum above 2 keV of SXP~59 is dominated by the non-thermal component from the accretion column, and can be phenomenologically fitted by a cut-off power-law component plus a hot blackbody emission. In this work, we do not find correlation between the parameters of the cut-off power-law model ($E_{\rm cut}$ and $\Gamma$) and the luminosity in the giant outburst of SXP~59 (Table \ref{spectra} and Figure \ref{nu_fit}). On the other hand, the behavior of hot blackbody emission is quite puzzling. If the source is at the sub-critical state, this component is generally considered to be from the base of accretion column, and contributes a large portion of X-ray flux  \citep[e.g. ][]{la09}. However, the prediction that the hot spot shrinks by a factor of $\sim 3$ during the outburst decay of SXP~59, conflicts with the constant normalization derived from the data (Figure \ref{nu_fit}). On the other side, theoretically, we can not receive the hot spot emissions directly at the high luminosity due to the accretion column. Nevertheless, the hot blackbody component is needed to fit the spectra of some luminous Be X-ray pulsars ($L_{\rm X} > 10^{36}-10^{39}$ erg~s$^{-1}$), e.g. GX~1+4 \citep{yoshida17}, EXO~2030+375 \citep{reig99}, SXP~59 (this work), in particular, Swift~J2043.6+6124 \citep{tao19}. The unexpected hot blackbody emission challenges the canonical accretion theories, which are mostly based on a pure dipole magnetic field. These observational results would indicate that either more physical spectral models are required to describe the spectra of luminous X-ray pulsars, or that the magnetic filed configuration deviates from a dipole field close to the NS surface.

\section*{Acknowledgements}

We thank the anonymous referee for the helpful comments, which significantly improve this work. SSW thanks Hua Feng and Lian Tao for valuable discussions. We acknowledge the use of public data from the High Energy Astrophysics Science Archive Research Center Online Service. This work is supported by the National Natural Science Foundation of China under grants 11673013, 11703014, 11503027, 11573023, U1838201, U1838104, and the Natural Science Foundation from Jiangsu Province of China (grant no. BK20171028).

%%%%%%%%%%%%%%%%%%%%%%%%%%%%%%%%%%%%%%%%%%%%%%%%%%%%%%%%%%%%%%%%%%%%%%%%%%%%%%%%%%%%%%%%%%%%%%%%%%%%%%%%%%%%%%%%%%%%%%%%%%%%%%%%%%%%%%%%%%%%%%%%%%%%%

%%%%%%%%%%%%%%%%%%%%%%%%%%%%%%%%%%%%%%%%%%%%%%%%%%%%%%%%%%%%%%%%
\end{document}